\documentclass[a4paper,fleqn,usenatbib,useAMS]{mnras}

\usepackage{tikz,hyperref}

\definecolor{lime}{HTML}{A6CE39}
\DeclareRobustCommand{\orcidicon}{
	\begin{tikzpicture}
	\draw[lime, fill=lime] (0,0) 
	circle [radius=0.13] 
	node[white] {{\fontfamily{qag}\selectfont \tiny ID}};
	\draw[white, fill=white] (-0.0625,0.095) 
	circle [radius=0.007];
	\end{tikzpicture}
	\hspace{-2mm}
}

\foreach \x in {A, ..., Z}{\expandafter\xdef\csname orcid\x\endcsname{\noexpand\href{https://orcid.org/\csname orcidauthor\x\endcsname}
			{\noexpand\orcidicon}}
}

\usepackage{graphicx}	
\usepackage{amsmath}	
\usepackage{amssymb}	
\usepackage{multicol}        
\usepackage{bm}		
\usepackage{pdflscape}	
\usepackage[encapsulated]{CJK}
\usepackage{ucs}
\usepackage[utf8x]{inputenc}
\usepackage{multirow}
\usepackage{booktabs,caption}
\usepackage[flushleft]{threeparttable}

\usepackage{rotating} 
\usepackage{lscape}

\usepackage[T1]{fontenc}
\usepackage{ae,aecompl}

\usepackage{newtxtext,newtxmath}

\title[Rings and arcs around evolved stars – III]
{Rings and arcs around evolved stars – III. Physical conditions of the ring-like structures in the planetary nebula IC\,4406 revealed by MUSE}
\author[Ramos-Larios et al.]{G.\,Ramos-Larios\thanks{E-mail: gerardo@astro.iam.udg.mx}$^{1}\orcidA$, J.~A.\,Toal\'{a}$^{2}\orcidD$, J.~B.\,Rodr\'{i}guez-Gonz\'{a}lez$^{2}\orcidC$,  M.~A.\,Guerrero$^{3}\orcidE$ \newauthor{and V.~M.~A.\,G\'{o}mez-Gonz\'{a}lez$^{2,4}\orcidB$}\\
$^{1}$Instituto de Astronom\'\i a y Meteorolog\'\i a, Dpto. de F\'\i sica, CUCEI, Univ.\ de Guadalajara, Av.\ Vallarta 2602, Arcos Vallarta, 44130 Guadalajara, Mexico\\
$^{2}$Instituto de Radioastronom\'{i}a y Astrof\'{i}sica (IRyA), UNAM Campus Morelia, Apartado postal 3-72, 58090 Morelia, Michoac\'{a}n, Mexico\\
$^{3}$Instituto de Astrof\'{i}sica de Andaluc\'{i}a, IAA-CSIC, Glorieta de la Astronom\'{i}a S/N, E-18008 Granada, Spain\\
$^{4}$Institute for Physics and Astronomy, Universit\"{a}t Potsdam, Karl-Liebknecht-Str. 24/25, D-14476 Potsdam, Germany
}

\pubyear{2022}

\begin{document}
\label{firstpage}
\pagerange{\pageref{firstpage}--\pageref{lastpage}}
\maketitle

\begin{abstract}
We present the analysis of Very Large Telescope (VLT) Multi Unit Spectroscopic Explorer (MUSE) observations of the planetary nebula (PN) IC\,4406. 
MUSE images in key emission lines are used to unveil the presence of at least 5 ring-like structures North and South of the main nebula of IC\,4406. 
MUSE spectra are extracted from the rings to unambiguously assess for the first time in a PN their physical conditions, electron density ($n_\mathrm{e}$) and temperature ($T_\mathrm{e}$). 
The rings are found to have similar $T_\mathrm{e}$ than the rim of the main nebula, but smaller $n_\mathrm{e}$. 
Ratios between different ionic species suggest that the rings of IC\,4406 have a lower ionization state than the main cavity, in contrast to what was suggested for the rings in NGC\,6543, the Cat's Eye Nebula.
\end{abstract}





\section{INTRODUCTION}
\label{sec:intro}

Ring-like structures have been detected around evolved low- and intermediate-mass stars (1~M$_\odot \lesssim M_\mathrm{i}<8~$M$_\odot$) in optical, IR and radio observations \citep[see, e.g.,][]{Mauron2006,Balick2012,Corradi2004,Phillips2009,RamosLarios2011,Cernicharo2015}. 
Such structures have been classified as (complete) rings, arcs, elliptical structures and spirals. 
They can be found in different stages of the evolution of those stars including asymptotic giant branch (AGB) stars, planetary nebulae (PNe), and sources in the intermediate proto-PN (pPN) evolutionary phase \citep[][hereinafter Paper~I]{RL2016}.
Throughout the present letter, we will refer to these structures simply as rings.

\begin{figure*}
\begin{center}
\includegraphics[width=0.98\linewidth]{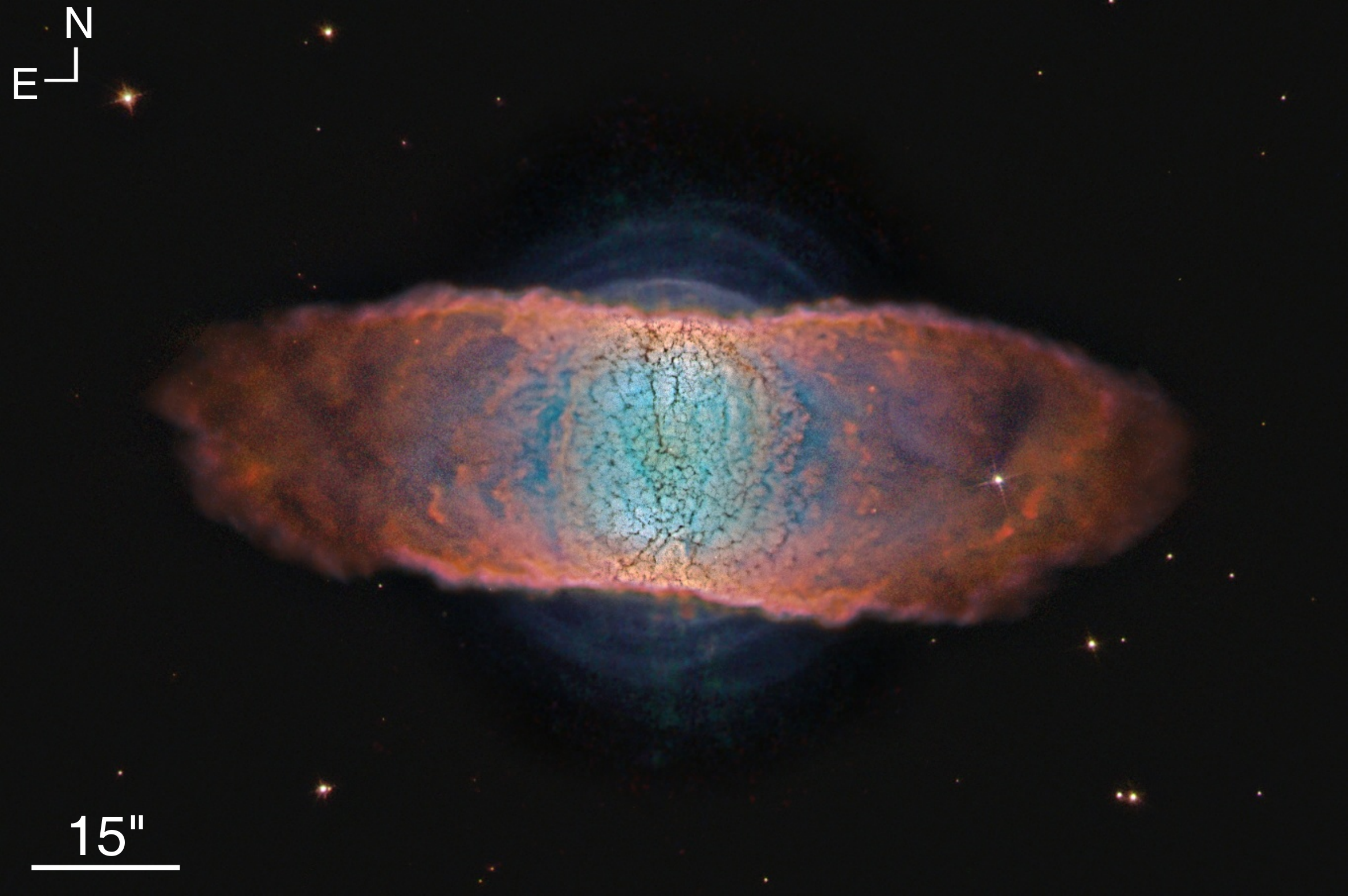}\\
\caption{Hybrid VLT MUSE+\emph{HST} WFPC2 color composite optical image of IC\,4406 ([O\,{\sc iii}] -- blue, H$\alpha$ -- green and [N\,{\sc ii}] -- red for {\it HST} and [O\,{\sc iii}] -- blue, [O\,{\sc i}] -- cyan, H$\alpha$ -- green, [N\,{\sc ii}] -- orange and [S\,{\sc ii}] -- red for MUSE). 
A series of ring-like structures North and South of the main nebula are clearly revealed thanks to the detailed MUSE-AOF data.
}
\label{fig:IC4406}
\end{center}
\end{figure*}

Different formation scenarios have been proposed in the literature, including the intrinsic variability of the wind of the AGB progenitor (e.g.\ Zijlstra et al.\ 2002), solar-like inversions of the stellar magnetic field (e.g.\ García-Segura et al.\ 2001), or the viscous coupling between the outflowing gas and dust components (e.g.\ Simis et al.\ 2001) to mention a few \citep[see, e.g.,][for a review; hereafter Paper~II]{Guerrero2020}.  
The formation scenario supported by the largest number of observational and numerical results rather involves the modulation of the AGB wind by the presence of a companion \citep[e.g,][]{Mastrodemos1999,Maercker2012}. 
Moreover, modern numerical simulations have shown that the variety of rings can be explain by the 3D spirals patterns produced by a binary depending on the viewing angle \citep[see][and references therein]{He2007,Kim2019,ElMellah2020}. 
Recent ALMA observations of AGB stars further suggest that sub-stellar companions might be able to produce similar structures \citep{Decin2020}.

In Paper~I we presented a search and characterization of rings around evolved low-and intermediate-mass stars using archival {\it Hubble Space Telescope (HST)} and {\it Spitzer Space Telescope} observations. 
Accounting for rings reported in the literature, we found that only 8\% of a sample of $\sim$650 objects exhibited any form of these structures. 
The reasons for such small occurrence may be twofold (possibly not excluding each other): 
the ring structures are less bright than the central cores of PNe and 
these structures are not present around all those evolved stars.

Detailed studies of rings around AGB stars have been relatively easy to achieve using sub-millimeter observations, proving to be one of the most powerful tools to study the velocity, density and temperature of spiral structures using their molecular emission. 
Subsequent modeling of the spirals give us information on the orbital parameters of the binary system \citep[see, e.g.,][]{Cernicharo2015,Kerschbaum2017,Doan2020}.

On the other hand, rings around pPNe and PNe seem to be more difficult to characterize. 
The expansion of the ionized rim that gives birth to the PN distorts the shape and even destroys the surrounding ring-like features (see Paper~II). 
Moreover, the main nebular shells are much brighter than the rings, requiring special  processing to be applied to the images (see Paper~I, and references therein). 
Studies of multi-epoch high-quality optical images obtained with the \emph{HST} as those presented by \citet{Balick2001,Balick2012} and Paper~II have successfully studied the expansion rates of the ring structures in the AGB C-rich star AFGL 3068, the pPNe CRL\,2688, and the PNe NGC\,6543, NGC\,7009 and NGC\,7027. 
Nevertheless, the  information on the physical properties of the ring-like structures, their electron temperature ($T_\mathrm{e}$) and density ($n_\mathrm{e}$), has not been properly investigated yet.

We present here the physical properties found in the ring system around a PN, namely IC\,4406, using the Integral Field Spectroscopic (IFS) technique that has been used to produce detailed studies of the two-dimensional variation of the excitation, ionization and chemical abundances of PNe \citep[e.g.]{Walsh2018,Walsh2020,MonrealIbero2020} and to unveil their spatio-kinematic structure \citep[e.g.,][]{Guerrero2021a}.  

IC\,4406 is a barrel-like bipolar PN with a bright waist and filamentary structures in the bipolar lobes (see Figure~\ref{fig:IC4406}). 
The waist can be interpreted as a toroidal structure surrounded by a dense torus of dust, whereas the bipolar lobes can be linked to CO ($J=2\rightarrow1)$ high-velocity flows along the nebular major axis \citep{Cox1991,Sahai1991}. 
In Paper~I we reported the presence of 3 rings in IC\,4406 immediately outside the dusty torus, two North of the bright waist and one fainter to its South. 
 
Here we use archival Very Large Telescope (VLT) Multi Unit Spectroscopic Explorer (MUSE) observations of IC\,4406 (see Fig.~\ref{fig:IC4406}), which provide unprecedented images and spectra of the rings around this PN to investigate their spatial distribution and physical conditions.

The observations used in this work are presented in Section~2, the images extracted from the IFS observations are described in Section~3, and the physical conditions of the rings are computed in Section~4.  
Discussion and conclusions are presented in Section~5.

\begin{figure*}
\begin{center}
\includegraphics[width=0.98\linewidth]{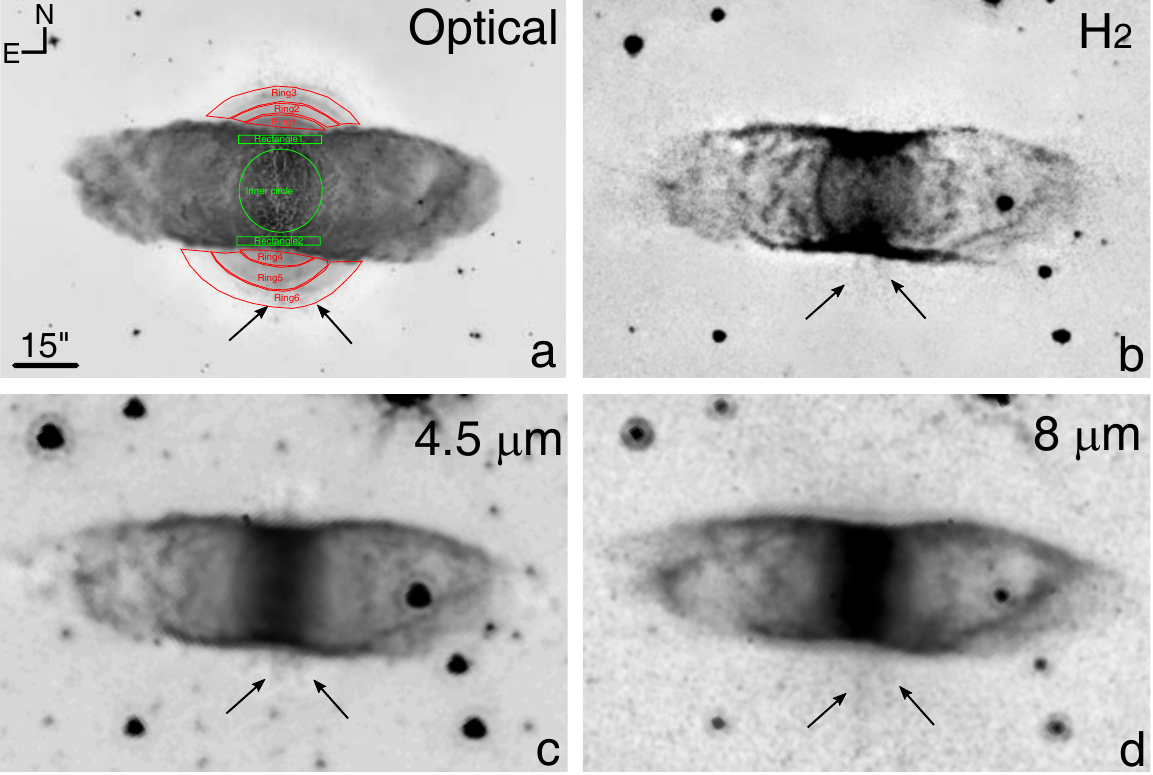}
\caption{
(a) Grey-scale representation of the hybrid image presented in Figure~\ref{fig:IC4406}. 
Spectra of the arc-like features are extracted from the red annulus sector apertures numerated from 1 to 6, and spectra of the main nebula from the green circular and rectangular regions (see text for details).
(b) CFHT near-IR H$_2$ 1--0 S(1) narrow-band image of IC\,4406.
(c) and (d) \emph{Spitzer} IRAC images of IC\,4406 in the 4.5 $\mu$m and 8 $\mu$m bands, respectively. 
The black arrows in all the images point to faint radial streaks described in Section~3.}
\label{fig:IC4406_2}
\end{center}
\end{figure*}

\section{OBSERVATIONS}

IC\,4406 was observed by the VLT MUSE on 2017 June 16 with a total exposure time of 180~s (Program~ID 60.A-9100; Obs.\,ID. 1755413). 
The observations cover the 4700--9350~\AA\, wavelength range with a spectral resolution of $R$=3014 and a pixel scale of 0.2~arcsec~pix$^{-1}$.
 
The Adaptive Optics Facility (AOF) instrument was used during the observations to compensate for atmospheric effects aiming at obtaining a sharper view. 
The AOF includes the Four Laser Guide Star Facility (4LGSF), whose beams make sodium (Na) atoms glow in the upper atmosphere imitating stars, and the Ground Atmospheric Layer Adaptive Corrector for Spectroscopic Imaging (GALACSI), which corrects the turbulence by means of a deformable secondary mirror. 
It is worth noting that the AOF uses a dichroic filter to block the light in the 5800 to 6000 \AA\ wavelength range, blocking the Na laser light from the detector.

The processed data were retrieved from the ESO Archive Science Portal\footnote{\url{http://archive.eso.org/scienceportal/home}}.

Images and spectra were extracted from the MUSE cube using the QFitsView analysis software\footnote{\url{https://www.mpe.mpg.de/~ott/QFitsView/}}. Preliminary spectra obtained by averaging over the brightest northern ring-like structures were obtained to identify the most prominent emission lines in order to produce images emphasizing the rings. 

To create high-quality images, we retrieved {\it HST} WFPC2 images from the Hubble Legacy Archive\footnote{\url{https://hla.stsci.edu/}} obtained on 2001 June 28 and 2002 January 19 (Prop.~ID 8726 and 9314). 
Images obtained in the 
F502N ($\lambda_{\rm c}=5012$ \AA, FWHM $=$ 26.9 \AA), 
F656N ($\lambda_{\rm c}=6564$ \AA, FWHM $=$ 21.5 \AA), and 
F658N ($\lambda_{\rm c}=6591$ \AA, FWHM $=$ 28.5 \AA) 
narrow-band filters were downloaded to produce images in the [O\,{\sc iii}] $\lambda$5007 \AA, H$\alpha$, and [N\,{\sc ii}] $\lambda$6584 \AA\ emission lines, respectively.

The observations for both epochs consisted of series of three exposures with exposure times of 600~s for the [O\,{\sc iii}] filter and 540~s for the H$\alpha$ and [N\,{\sc ii}] filters. 
We also use in our composition the image downloaded from the website \url{https://commons.wikimedia.org/wiki/File:IC_4406_"Retina".png}.

Figure~\ref{fig:IC4406} presents a color-composite, hybrid MUSE+{\it HST} nebular image of IC\,4406. 
The hybrid composition is actually the result of layering two separate frames of the same object obtained using different instruments (in this case, the {\it HST} WFPC2 and VLT MUSE). 
This technique, extensively used by our group \citep[see, e.g.,][and references therein]{Guerrero2021a,Guerrero2021b,Sabin2021,Toala2021,RamosLarios2018}, registers the lower-resolution VLT MUSE image to match the higher-resolution \emph{HST} WFPC2 image.  
In particular, the {\it HST} images were combined with narrow-band images extracted from the MUSE data cube in the emission lines of [O\,{\sc iii}] $\lambda5007$ \AA, [O\,{\sc i}] $\lambda6300$ \AA, H$\alpha$, [N\,{\sc ii}] $\lambda6584$ \AA, and [S\,{\sc ii}] $\lambda6731$ \AA.
This procedure enhances the small-scale structures and fine details features of the main nebula present in the {\it HST} image and emphasizes the outer emission mainly from the rings and high- and low-ionization structures present in the MUSE data.
 
A gray scale version of this image is presented in Figure~\ref{fig:IC4406_2}-{\it a}.

For comparison and discussion we also obtained archival near-IR H$_2$ images from the 3.6\,m Canada-France-Hawaii Telescope (CFHT) IR camera (CFHTIR), which uses a 1024$\times$1024 Rockwell HAWAII \#142 detector with a pixel scale of 0.2 arcsec~pix$^{-1}$. 
A series of ten exposures 180~s long were acquired on 2004 May 4 (Program ID 04AT02, P.I. Sun Kwok) through the H$_2$ 1$-$0 S(1) ($\lambda_{c}$=2.122 $\mu$m) filter. 
The images were combined and stacked for the final composition shown in Figure~\ref{fig:IC4406_2}-{\it b}.

Finally, we also retrieved {\it Spitzer} IRAC data from the Spitzer Heritage Archive \footnote{\url{https://sha.ipac.caltech.edu/applications/Spitzer/SHA/}}.
The observations were performed on 2004 March 5 under the Program ID 68 (PI: G.\,Fazio).
{\it Spitzer} IR images of IC\,4406 are also shown in Figure~\ref{fig:IC4406_2}-{\it c} and {\it d}.

\begin{figure}
\begin{center}
\includegraphics[width=\linewidth]{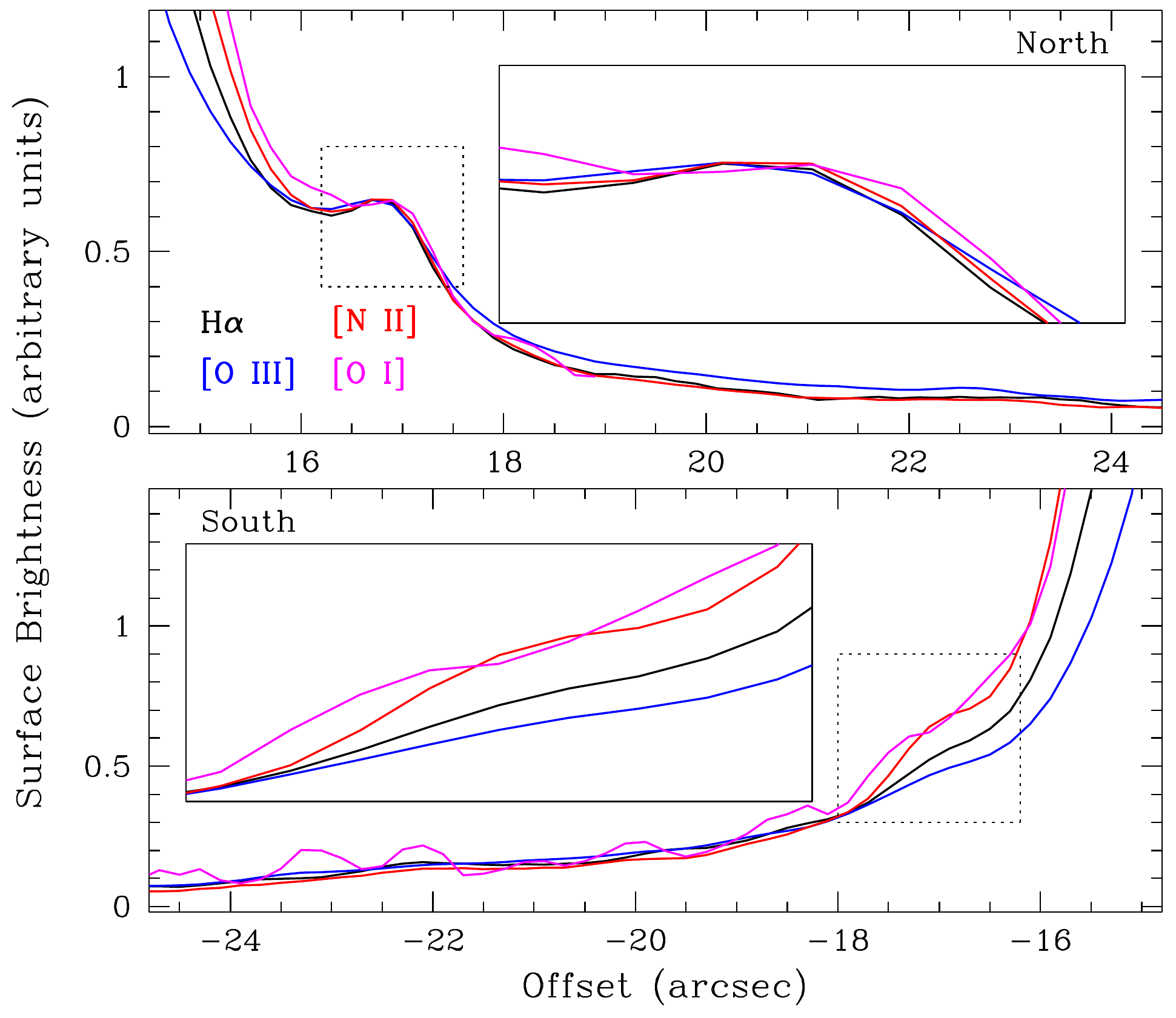}
\caption{
Surface brightness spatial profiles of the rings of IC\,4406 along PA 358$^{\circ}$ at the North (top) and PA 178$^{\circ}$ at the South (bottom) of its main nebular shell in the H$\alpha$ (black), [O~{\sc iii}] $\lambda$5007 \AA\ (blue), [N~{\sc ii}] $\lambda$6584 \AA\ (red), and [O~{\sc i}] $\lambda$6300 \AA\ (magenta) emission lines.
The dotted inbox is zoomed into the solid box to emphasize the spatial offsets between different emission lines that reveal the rings ionization structure.
}
\label{fig:mar}
\end{center}
\end{figure}

\section{Imagery}

The main (bright) body of IC\,4406, the elongated barrel-like structure, is detected in a wealth of emission lines. 
This is not the case for the rings, however, which are best detected in the [O\,{\sc iii}] $\lambda$5007 \AA\ emission line image. 
The hybrid image presented in Figure~\ref{fig:IC4406} suggests the presence of a series of at least 5 arcs northern and southern of IC\,4406 extending to distances up to 30~arcsec from the central star, with the innermost northern arc being the brightest one.  
These images demonstrate that the ring-like structures do not have a coherent shape, i.e.\ they do not seem to correspond to sections of complete circular structures, neither spirals. 
They are rather consistent with interlaced arcs projected onto the plane of the sky disrupted along the E-W direction by the expansion of the PN.  

The optical and IR images of IC\,4406 presented in Figure~\ref{fig:IC4406} and \ref{fig:IC4406_2} reveal the presence of radial streaks protruding from the inner cavity of IC\,4406 and pointing toward the position of the central star (black arrows in these figures). 
These rays can also be hinted in the {\it Spitzer} 4.5~$\mu$m image, as well as in the H$_2$ image, although weaker and less extended (see other panels in Fig.~\ref{fig:IC4406_2}). 
These structures can be caused by clumps in the main nebular shell that would produce an azimuthally non-uniform photoionization of the outermost nebular regions.
Indeed, the optical grey-scale image presented in Figure~\ref{fig:IC4406_2}-{\it a} shows that these radial streaks are detected farther than the rings. 
This implies the azimuthally variable ionization of the arcs, which, according to Paper~II, induces azimuthally non-uniform variations in the pressure of the gas that can be expected to affect the shape of the arcs, leading ultimately to their disruption.

\begin{figure*}
\begin{center}
\includegraphics[width=\linewidth]{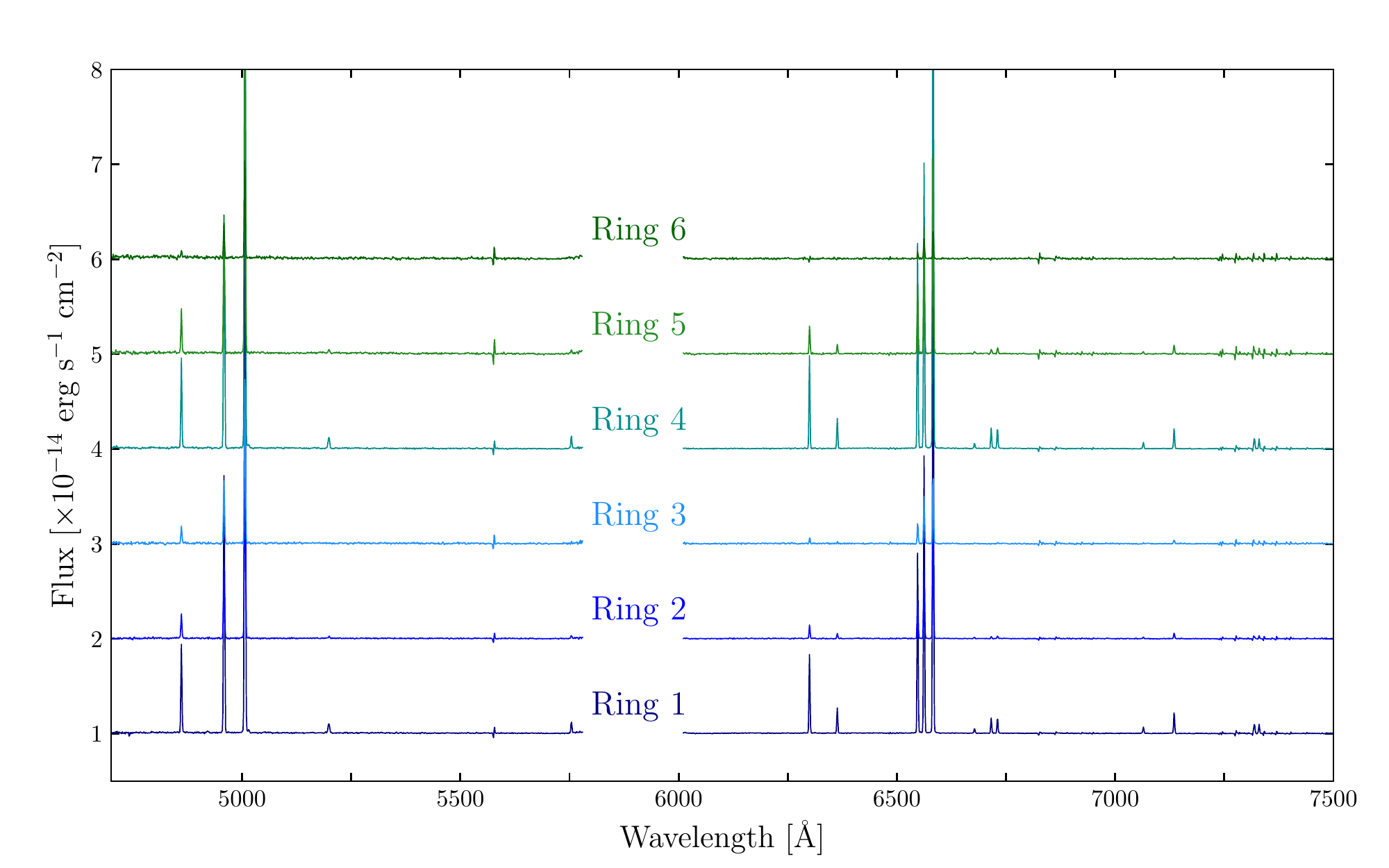}
\caption{
VLT MUSE spectra in the 4700--7500~\AA\, wavelength range of  ring-like features in IC\,4406 extracted from the six red apertures overlaid in Figure~\ref{fig:IC4406_2}-{\it a}.  
Note that the AOF dichroic filter blocks the light in the 5800 to 6000 \AA\ wavelength range. 
}
\label{fig:spec}
\end{center}
\end{figure*}

\begin{table*}
\footnotesize	
\setlength{\tabcolsep}{0.35\tabcolsep}
\caption{Dereddened line fluxes relative to H$\beta$=100 for 
the combined spectra extracted from the ring-like of IC\,4406.}
\begin{center}
\begin{tabular}{ccccccccccc}
\hline
$\lambda_{0}$ & line            & Ring\,1          & Ring\,2       & Ring\,3        & Ring\,4      & Ring\,5       & Ring\,6         & Inner C.       & Rec.\,1      & Rec.\,2      \\
(\AA)         &                 &                  &               &                &              &               &                 &                &              &              \\
\hline
4711.4 & [Ar{\,\scshape iv}]    & \dots            & \dots         & \dots          &  1.3$\pm$0.4 & \dots         &                 & 2.3$\pm$0.1    & 0.8$\pm$0.1   & 0.8$\pm$0.1 \\
4740.2 & [Ar{\,\scshape iv}]    & \dots            & \dots         & \dots          &  \dots       & \dots         &                 & 1.4$\pm$0.1    & 0.3$\pm$0.1   & 0.2$\pm$0.1 \\
4861.4 & H$\beta$               & 100              & 100           & 100            & 100          & 100           & 100             & 100            & 100           & 100         \\    
4921.9 & He{\,\scshape i}       & 2.7$\pm$0.7      & \dots         & \dots          & \dots        & \dots         & \dots           & 1.3$\pm$0.1    & 1.6$\pm$0.1   & 1.7$\pm$0.1 \\  
4958.9 & [O{\,\scshape iii}]    & 284.2$\pm$3.5 & 317$\pm$19 &  255$\pm$27 & 251$\pm$6 & 305$\pm$11& 450$\pm$50  & 381$\pm$0.3    & 282$\pm$0.4   & 267.2$\pm$0.3\\  
5006.8 & [O{\,\scshape iii}]    & 831$\pm$6     & 907$\pm$16 & 1028$\pm$37 & 734$\pm$6 & 884$\pm$14& 1370$\pm$50 & 1120.0$\pm$0.2 & 822.2$\pm$0.6 & 789.0$\pm$0.5\\
5015.7 & {[He\,{\sc i}]}        & 5.8$\pm$0.8   & \dots         & \dots          & 5.4$\pm$0.7  & \dots         & \dots           & 3.8$\pm$0.1    & 3.7$\pm$0.1   & 3.7$\pm$0.1  \\
5200.0 & {[N{\,\scshape i}]}    & 15.6$\pm$0.8  & 9.8$\pm$2.1   & \dots          & 19.8$\pm$1.2 & 13.4$\pm$2.3  & \dots           & 4.7$\pm$0.1    & 12.8$\pm$0.1  & 13.7$\pm$0.1 \\
5517.7 & {[Cl{\,\scshape iii}]} & 1.1$\pm$0.4   & \dots         & \dots          & \dots        & \dots         & \dots           & 0.6$\pm$0.1    & 0.8$\pm$0.1   & 0.8$\pm$0.1  \\
5537.9 & {[Cl{\,\scshape iii}]} & 0.7$\pm$0.3   & \dots         & \dots          & \dots        & \dots         & \dots           & 0.5$\pm$0.1    & 0.7$\pm$0.1   & 0.7$\pm$0.1  \\
5577.3 & {[O\,{\sc i}]}         & 4.3$\pm$1.5   & 14.7$\pm$2.9  & 38.7$\pm$3.8   & 5.8$\pm$0.8  & 22.0$\pm$5.3  & \dots           & 0.4$\pm$0.1    & 1.0$\pm$0.1   & 1.0$\pm$0.1  \\
5754.6 & {[N{\,\scshape ii}]}   & 11.9$\pm$0.6  & 13.0$\pm$1.1  & \dots          & 14.1$\pm$0.7 & 7.9$\pm$1.5   & \dots           & 5.7$\pm$0.1    & 16.0$\pm$0.1  & 16.4$\pm$0.2 \\
6300.3 & {[O{\,\scshape i}]}    & 82.1$\pm$1.6  & 51.3$\pm$3.8  & 32$\pm$7   & 94.3$\pm$2.4 & 60.2$\pm$8.2  & \dots           & 22.0$\pm$0.1   & 69.3$\pm$0.1  & 72.2$\pm$0.1 \\
6312.1 & {[S{\,\scshape iii}]}  & \dots         & \dots         & \dots          &  \dots       & \dots         & \dots           & 0.7$\pm$0.1    & 0.8$\pm$0.1   & 0.8$\pm$0.1  \\
6363.8 & {[O{\,\scshape i}]}    & 270$\pm$1.3   & 19.68$\pm$2.5 & 11$\pm$6   & 31.1$\pm$1.0 & 18.1$\pm$2.7  & \dots           & 7.2$\pm$0.1    & 22.4$\pm$0.1  & 23.3$\pm$0.1 \\
6548.1 & {[N{\,\scshape ii}]}   & 187.9$\pm$2.1 & 150$\pm$6   & 113$\pm$10 & 208.4$\pm$3.4& 143$\pm$10 & 64$\pm$11   & 9.8$\pm$0.1    & 266.7$\pm$0.2 & 272.7$\pm$0.3 \\ 
6562.8 & H$\alpha$              & 287.4$\pm$2.8    & 262$\pm$7 & 258$\pm$12 & 289.6$\pm$3.4& 263$\pm$14& 258$\pm$21  & 270$\pm$0.1    & 292.6$\pm$0.2 & 289.8$\pm$0.3 \\
6583.5 & {[N{\,\scshape ii}]}   & 575.9$\pm$2.3    & 467$\pm$7 & 374$\pm$12 & 637.7$\pm$3.9& 450$\pm$6 & 347$\pm$12  & 289$\pm$0.1    & 807.4$\pm$0.4 & 828.1$\pm$0.4 \\
6678.2 & {He{\,\scshape i}}     & 6.2$\pm$0.6      & 7.1$\pm$1.9   & \dots          & 5.5$\pm$0.5  & 5.5$\pm$1.2   & \dots           & 3.7$\pm$0.1    & 4.9$\pm$0.1   & 4.9$\pm$0.1   \\
6716.4 & {[S{\,\scshape ii}]}   & 16.2$\pm$0.8     & 10.8$\pm$2.1  & \dots          & 20.8$\pm$0.7 & 10.4$\pm$1.3  & \dots           & 5.8$\pm$0.1    & 19.9$\pm$0.1  & 23.9$\pm$0.1  \\
6730.8 & {[S{\,\scshape ii}]}   & 16.6$\pm$0.6     & 11.1$\pm$2.1  & 5.2$\pm$2.5    & 21.2$\pm$0.7 & 11.9$\pm$1.3  & \dots           & 6.4$\pm$0.1    & 22.1$\pm$0.1  & 25.9$\pm$0.1  \\
7065.2 & {He{\,\scshape i}}     & 6.8$\pm$0.6      & 5.6$\pm$1.6   & 4.9$\pm$2.6    & 7.0$\pm$0.5  & 5.4$\pm$1.9   & \dots           & 3.3$\pm$0.1    & 5.7$\pm$0.1   & 5.6$\pm$0.1  \\
7135.8 & {[Ar{\,\scshape iii}]} & 22.0$\pm$0.8     & 22.8$\pm$2.5  & 22$\pm$4   & 21.3$\pm$0.9 & 21.6$\pm$2.3  & 36$\pm$12   & 22.3$\pm$0.1   & 29.3$\pm$0.1  & 28.8$\pm$0.1  \\
7319.0 & {[O{\,\scshape ii}]}   & 13.4$\pm$2.0     & 14$\pm$11 & 16.6$\pm$3.5   & 13.4$\pm$1.5 & 14$\pm$6  & 53$\pm$17   & 5.8$\pm$0.1    & 16.7$\pm$0.2  & 16.3$\pm$0.2  \\
7330.2 & {[O{\,\scshape ii]}}   & 9.8$\pm$2.6      & 9.4$\pm$4.1   & 11.8$\pm$3.3   & 9.3$\pm$1.7  & 11$\pm$5  & 27$\pm$17   & 4.7$\pm$0.1    & 13.4$\pm$0.1  & 13.1$\pm$0.3  \\
7751.1 & {[Ar{\,\scshape iii}]} & 6.9$\pm$1.6      & 12$\pm$12 & \dots          & 6.6$\pm$3.7  & 16$\pm$6  & 72$\pm$37   & 5.2$\pm$0.1    & 7.0$\pm$0.1   & 6.8$\pm$0.2   \\
9068.6 & {[S{\,\scshape iii}]}  & 8.0$\pm$1.7      & 7.2$\pm$3.4   & \dots          & 8.8$\pm$3.5  & 13$\pm$7  & 23$\pm$16   & 9.0$\pm$0.1    & 10.5$\pm$0.2  & 11.3$\pm$0.3  \\
\hline
log($F$(H$\beta$)) & [erg~s$^{-1}$~cm$^{-2}$] & $-$13.54$\pm$0.01 & $-$14.11$\pm$0.03 & $-$14.26$\pm$0.05 & $-$13.53$\pm$0.01 & $-$13.84$\pm$0.04 & $-$14.65$\pm$0.08 & $-$10.76$\pm$0.01 & $-$11.87$\pm$0.01 & $-$11.89$\pm$0.01 \\
$c$(H$\beta$)      &                          & 0.06$\pm$0.11     & 0.00$\pm$0.10 & 0.00$\pm$0.10 & 0.10$\pm$0.01 & 0.00$\pm$0.10 & 0.00$\pm$0.20 & 0.27$\pm$0.01 & 0.36$\pm$0.01 & 0.35$\pm$0.01 \\
\hline
log($T_\mathrm{e}$([N\,{\sc ii}])) & [K]         & 4.06$\pm$0.01 & 4.14$\pm$0.03 & \dots & 4.09$\pm$0.01 & 4.04$\pm$0.04 & \dots & 4.12$\pm$0.01& 4.06$\pm$0.01 & 4.07$\pm$0.01\\
log($T_\mathrm{e}$([S\,{\sc iii}])) & [K]         & \dots  & \dots & \dots & \dots & \dots & \dots & 4.06$\pm$0.02 & 4.05$\pm$0.02 & 4.04$\pm$0.02\\
log($n_\mathrm{e}$([S\,{\sc ii}])) & [cm$^{-3}$] & 2.85$\pm$0.16 & 2.85$\pm$0.37 & \dots & 2.84$\pm$0.12 & 3.03$\pm$0.42 & \dots & 2.98$\pm$0.06 & 2.99$\pm$0.02 & 2.95$\pm$0.01 \\
log($n_\mathrm{e}$([Cl\,{\sc iii}])) & [cm$^{-3}$] & \dots & \dots & \dots & \dots & \dots & \dots & 3.03$\pm$0.30 & 3.15$\pm$0.20 & 3.15$\pm$0.20 \\
\hline
\end{tabular}\\
\label{tab:obs} 
\end{center}
\end{table*}

To peer into the ionized structure of the rings, surface brightness profiles of emission lines that map the rings have been extracted along directions of interest. 
Figure~\ref{fig:mar} shows the H$\alpha$, [N\,{\sc ii}] $\lambda$6584 \AA, [O\,{\sc i}] $\lambda$6300 \AA, and [O\,{\sc iii}] $\lambda$5007 \AA\ profiles extracted from the northern (PA=358$^\circ$) and southern (PA=178$^\circ$) rings. 
Whereas the H$\alpha$, [N\,{\sc ii}], and [O\,{\sc iii}] lines exhibit almost identical profiles, peaking at the same positions, this is not the case for the [O\,{\sc i}] emission. The peak of the latter is displaced outwards, with the  [O\,{\sc i}] emission peaks of the brightest northern and southern ring shifted by $\simeq$0\farcs2 and $\simeq$0\farcs3, respectively, with respect to those of the other emission lines.
 This unveils a notable ionization stratification of the rings, most likely associated with a photoionization front, as shocks in a low-density medium are expected to produce inverted ionization structures instead (e.g., IC\,4634, Guerrero et al.\ 2008).

\section{Physical conditions of the rings}

We extracted several spectra from a number of apertures probing different regions within IC\,4406.  
Six of these spectra correspond to the rings at the northern and southern regions from IC\,4406 (red annulus sectors in Fig.~\ref{fig:IC4406_2}-{\it a}). 
These spectra are presented in Figure~\ref{fig:spec}.
 
Spectra from the outermost rings were also extracted, but their low signal-to-noise ratio ($S/N$) hampered any further spectral analysis.
The extracted spectra were corrected for extinction by using the $c$(H$\beta$) value estimated for each region from the Balmer decrement method corresponding to a case B photoionized nebula of $T_\mathrm{e}$ = 10,000 K and $n_\mathrm{e}$ = 1000~cm$^{-3}$ \citep{Osterbrock2006} and the reddening curve of \citet{Cardelli1989} with $R_\mathrm{V}$=3.1.

For control and comparison, we also extracted three additional spectra, one from a circular aperture at the center of IC\,4406 representative of high-excitation regions and two rectangular apertures on the northern and southern rim of IC\,4406 representative of low-excitation regions. 
The extraction regions in green corresponds to a circular aperture with radius 9.5~arcsec as shown in Figure~\ref{fig:IC4406_2}-{\it a}, which will now be referred as Inner Circle, and two rectangular regions of 19$\times$1.8~arcsec in size referred as Rectangle 1 and 2, respectively.
The spectral analysis has been performed using {\sc iraf} \citep{Tody1993} standard routines.

Table~\ref{tab:obs} presents the complete list of lines detected from the spectra of IC\,4406 extracted from different regions.
Dereddened intensities relative to an arbitrary value 100 for H$\beta$ are also listed in this table. Table~\ref{tab:obs} corroborates previous statements indicating that the [O\,{\sc iii}] 5007~\AA\ emission line dominates in the rings (see Paper II, and references therein), followed by the [N\,{\sc ii}] 6583~\AA\ emission line.
The available emission lines allowed us to compute $T_\mathrm{e}$ using the [N\,{\sc ii}] and [S~{\sc iii}] auroral-to-nebular line ratio\footnote{
The value of $T_\mathrm{e}$([O\,{\sc iii}]) could not be computed as the spectral range of the MUSE data does not include the [O III] 4363~\AA\ auroral emission line.
} 
and $n_\mathrm{e}$ using the [S\,{\sc ii}] $\lambda\lambda$6717,6730 \AA\ 
and [Cl~{\sc iii}] $\lambda\lambda$5518,5538 \AA\ 
doublet line ratios. 
This calculation has been performed using the {\sc iraf} task {\it temden}. 
The estimated values of $T_\mathrm{e}$ and $n_\mathrm{e}$ are listed in the bottom section of Table~\ref{tab:obs}.

The spectra extracted from the inner rings (Ring 1 and 4) resulted in $n_\mathrm{e}\approx700$~cm$^{-3}$ and $T_\mathrm{e}\approx12,000$ K. 
The spectra of the most external rings (not shown here) did not detect the necessary emission lines to calculate $n_\mathrm{e}$ and $T_\mathrm{e}$. 
On the other hand, the spectra obtained for the Rectangle 1 and 2 extractions exhibit a large number of emission lines with higher $S/N$ (see last two columns of Table~\ref{tab:obs}). 
For these spectra, the diagnostic-sensitive line ratios implied $n_\mathrm{e}$ in the range 900 to 1,000 cm$^{-3}$ and $T_\mathrm{e}$ in the range 11,500 to 13,100 K.

\section{Discussion and conclusions}

\begin{table*}
\footnotesize	
\setlength{\tabcolsep}{0.8\tabcolsep}
\caption{Ionization fractions of different regions in IC\,4406.}
\begin{center}
\begin{tabular}{cccccccc}
\hline
                & Ring1             & Ring2             & Ring4             & Ring5             & Inner C           & Rec.~1.          &    Rec.~2       \\
\hline
N$^{+}$/H$^{+}$ & 0.0077$\pm$0.0004 & 0.0042$\pm$0.0004 & 0.0073$\pm$0.0003 & 0.0066$\pm$0.0013 & 0.0028$\pm$0.0001 & 0.0107$\pm$0.0005 & 0.0105$\pm$0.0005 \\
O$^{0}$/H$^{+}$ & 0.0104$\pm$0.0005 & 0.0036$\pm$0.0003 & 0.0096$\pm$0.0004 & 0.0089$\pm$0.0015 & 0.0018$\pm$0.0001 & 0.0087$\pm$0.0006 & 0.0085$\pm$0.0006 \\ 
O$^{+}$/H$^{+}$ & 0.0085$\pm$0.0008 & 0.0034$\pm$0.0006 & 0.0060$\pm$0.0006 & 0.0098$\pm$0.0063 & 0.0017$\pm$0.0003 & 0.0096$\pm$0.0012 & 0.0087$\pm$0.0009 \\ 
O$^{++}$/H$^{+}$& 0.0195$\pm$0.0011 & 0.0128$\pm$0.0011 & 0.0141$\pm$0.0008 & 0.0239$\pm$0.0055 & 0.0178$\pm$0.0010 & 0.0192$\pm$0.0012 & 0.0173$\pm$0.0011 \\
\hline 
N$^+$/O$^+$    & 0.82$\pm$0.14      & 1.23$\pm$0.40     & 1.22$\pm$0.19     & 0.67$\pm$1.58     & 1.65$\pm$0.43     & 1.12$\pm$0.22     & 1.20$\pm$0.20 \\
O$^{++}$/O$^0$ & 1.87$\pm$0.20      & 3.55$\pm$0.65     & 1.46$\pm$0.15     & 2.68$\pm$1.28     & 9.90$\pm$1.17     & 2.20$\pm$0.31     & 2.04$\pm$0.30 \\
O$^{++}$/O$^+$ & 2.29$\pm$0.38      & 3.76$\pm$1.20     & 2.35$\pm$0.40     & 2.43$\pm$5.96     & 10.47$\pm$2.95    & 2.00$\pm$0.43     & 2.00$\pm$0.37 \\ 
O$^+$/O$^0$    & 0.81$\pm$0.12      & 0.94$\pm$0.27     & 0.62$\pm$0.10     & 1.10$\pm$1.07     & 0.95$\pm$0.23     & 1.10$\pm$0.23     & 1.02$\pm$0.19 \\ 
\hline
\end{tabular}\\
\label{tab:ratios} 
\end{center}
\end{table*}

The detection of ring-like structures in PNe has been a difficult task in the past (see Paper~I).  
The determination of their physical conditions has been even more challenging as the high-quality spectroscopic data required to detect the temperature and density sensitive emission lines of the elusive ring emission were lacking. 
IFS observations offer the means to obtain direct determinations of the physical conditions $n_\mathrm{e}$ and $T_\mathrm{e}$ of the ring-like structures in any PN using spectra. 
The analysis of the spectra extracted from the MUSE data indicates that $n_\mathrm{e}$ decreases from 900--1,000 cm$^{-3}$ in the main cavity and rim of IC\,4406 to $\sim$700~cm$^{-3}$ for the first outer rings where the density determination is more reliable. 
Meanwhile $T_\mathrm{e}$ is virtually the same for the rim of IC\,4406 and the first outer rings (see Table~\ref{tab:obs}), with some evidence of larger temperatures for the innermost regions.

We can use the emission line intensities, $n_\mathrm{e}$, and $T_\mathrm{e}$ values listed in Table~\ref{tab:obs} to assess possible ionization differences between the main nebular shell of IC\,4406 and its ring-like structures. 

\begin{figure*}
\begin{center}
\includegraphics[width=0.98\linewidth]{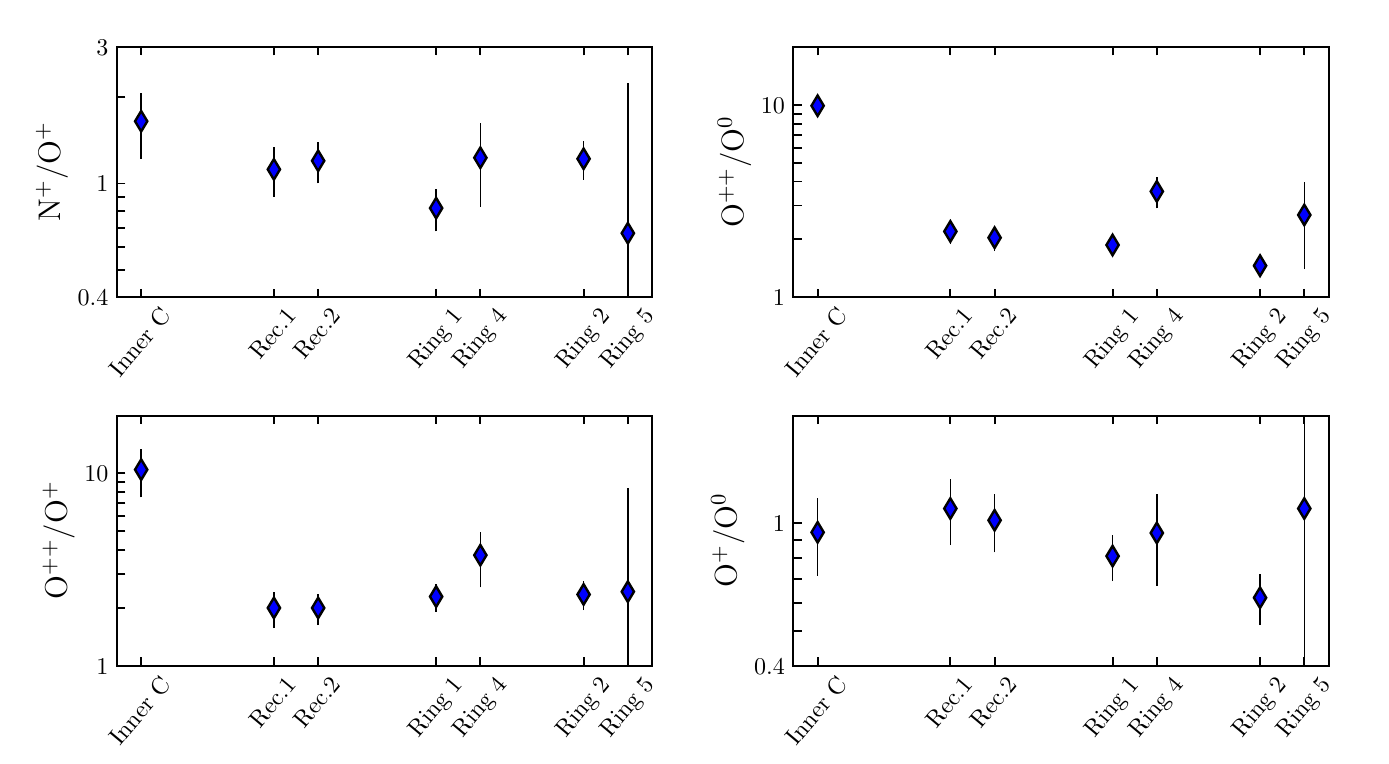}
\caption{
Ionic ratios obtained for different regions in IC\,4406. These results are also listed in Table~\ref{tab:ratios}. 
The dots are distributed according to the distance of each region to the central star. 
}
\label{fig:fig_ratios}
\end{center}
\end{figure*}

Using the expressions listed in table~3 of \citet{Alexander1997}, we calculated the ionic ratios listed in Table~\ref{tab:ratios} and presented in Figure~\ref{fig:fig_ratios} for a better inspection. 
Compared to the inner regions of the main nebula, the nebula edge and the ring-like features are characterized by a lower ionization stage, with lower values of the O$^{++}$/O$^0$ and O$^{++}$/O$^+$ ratios (Fig.~\ref{fig:fig_ratios}). 
Meanwhile the O$^{+}$/O$^0$ ratio is notably flat across the different nebular structures. 
The N$^+$/O$^+$ ionic ratios in the ring-like features are also similar to those of the nebular rim, whereas it is consistent for the innermost nebula given the large uncertainty for this value.   
If this ratio is assumed to represent the N/O abundance ratio, as usually adopted \citep{Kingsburgh1994}, this would indicate similar N/O abundance ratio for the outer rings and main nebula.

Similar calculations have only been attempted for NGC\,6543, the Cat's Eye Nebula \citep{Balick2001}. 
These authors used narrow-band {\it HST} images and spectra of the inner core of NGC\,6543 to suggest that the rings shared similar properties as that of the main nebula, which is in contrast with the results of the analysis of the MUSE data of IC\,4406 here presented. Further analysis of high-quality spectra of ring-like structures are most needed to assess their physical structure and ionization stage.

Finally, we note the spatial shift of the peak intensity of the [O~{\sc i}] emission with respect to the other emission lines. 
This reveals a notable ionization stratification within the rings.

\section*{Acknowledgments}

GR-L acknowledges support from CONACyT grant 263373 and PRODEP
(Mexico).  VMAGG acknowledges support from the Programa de Becas
posdoctorales of the Direcci\'{o}n General de Asuntos del Personal
Acad\'{e}mico (DGAPA) of the Universidad Nacional Aut\'{o}noma de
M\'{e}xico (UNAM, Mexico).  VMAGG and JAT acknowledge funding by DGAPA
UNAM PAPIIT project IA100720. JAT also acknowledges support from the
Marcos Moshinsky Fundation (Mexico). JBR-G thanks CONACyT for a student scholarship. 
MAG acknowledges support of the Spanish Ministerio de Ciencia, Innovaci\'{o}n y Universidades grant
PGC2018-102184-B-I00, co-funded by FEDER funds.
Based on observations made with the NASA/ESA Hubble Space Telescope, and obtained from the Hubble Legacy Archive, which is a collaboration between the Space Telescope Science Institute (STScI/NASA), the Space Telescope European Coordinating Facility (ST-ECF/ESA) and the Canadian Astronomy Data Centre (CADC/NRC/CSA).
Based on observations collected at the European Organization for Astronomical Research in the Southern Hemisphere under ESO programme 60.A-9100. Based on observations obtained at the Canada-France-Hawaii Telescope (CFHT). This work has make
extensive use of the NASA’s Astrophysics Data System.

\section*{Data availability}

The data underlying this work are available on request to the first author. The processed observations files can be found in the ESO Archive Science Portal.


\end{document}